\documentclass[aps,pra,reprint,amsmath,amssymb,groupedaddress,showpacs,nobibnotes]{revtex4-1}
\usepackage{graphicx}
\usepackage{dcolumn}
\usepackage{bm}
\usepackage{hyperref}

\begin{document}
\title{Observing photonic de Broglie waves without the NOON state}

\author{Osung Kwon }
\affiliation{Department of Physics, Pohang University of Science and
Technology (POSTECH), Pohang, 790-784, Korea}

\author{Young-Sik Ra }
\affiliation{Department of Physics, Pohang University of Science and
Technology (POSTECH), Pohang, 790-784, Korea}

\author{Yoon-Ho Kim}
\email{yoonho72@gmail.com} 
\affiliation{Department of Physics,
Pohang University of Science and Technology (POSTECH), Pohang,
790-784, Korea}

\date{\today}

\begin{abstract}
The photonic de Broglie wave, in which an ensemble of $N$ identical photons with wavelength $\lambda$ reveals $\lambda/N$ interference fringes, has been known to be a unique feature exhibited by the photon number-path entangled state or the NOON state. Here, we report the observation of the photonic de Broglie
wave for a pair of photons, generated by spontaneous parametric down-conversion, that are not
photon number-path entangled. We also show that the photonic de Broglie wave can even be observed for a pair of photons that are completely separable (i.e., no entanglement in all degrees of freedom) and distinguishable. The experimental and theoretical results suggest that the photonic de Broglie wave is, in fact, not related to the entanglement of the photons, rather it is related to the indistinguishable pathways established by the measurement scheme.
\end{abstract}


\maketitle

\section{Introduction}

The nature of multipartite quantum entanglement is often manifested
in quantum interference experiments. For example, in the case of
entangled photon states generated by spontaneous parametric
down-conversion (SPDC), quantum interference is observed in
coincidence counts between two detectors, each individually
exhibiting no interference fringes \cite{rarity,ou,brendel}.

One notable example of photonic quantum interference is the photonic de Broglie wave in which an ensemble of $N$ identical photons with wavelength $\lambda$ exhibits $\lambda/N$ interference fringes \cite{jacob}. The photonic de Broglie wavelength $\lambda/N$ can be observed at the $N$-photon detector placed at an output port of an interferometer if the beam splitters that make up the interferometer do not randomly split $N$ photons. The quantum state of the photons in the interferometer is then the photon number-path entangled state or the NOON state
\begin{equation}
|\psi\rangle = (|N\rangle_1|0\rangle_2 + |0\rangle_1|N\rangle_2)/\sqrt{2},
\end{equation}
where the subscripts refer to the two interferometric paths. For this reason, the photonic de Broglie wave has been considered to be a unique feature exhibited by  the NOON state and essential for quantum imaging and quantum metrology \cite{boto,dangelo,kapale}. Experimentally, photonic de Broglie waves up to $N=4$ have been observed with corresponding NOON states \cite{fon,eda,walther,mitchell,shige}.

Note, however, that $\lambda/N$ modulations in the coincidence rate
among $N$ detectors may not necessarily be of quantum origin. For
instance, $\lambda/N$ modulation in coincidences
among $N$ detectors reported in Ref.~\cite{resch}, with each detector placed at an output port of a multi-path
interferometer, is a classical effect since the coincidence
modulation is a direct result of modulations (with different phases)
observed at individual detectors. Also, classical
thermal light may exhibit sub-wavelength interference fringes in
coincidences but at the reduced visibility consistent with classical
states  \cite{ferri,xiong}. Thus, the reduced-period fringe itself need not be of quantum origin. It is, however, important to point out that $N$-th order quantum interference, such as quantum optical $\lambda/N$ modulations due to the photonic de Broglie wave,  must exhibit high visibility (up to 100\% in principle) in the absence of any lower-order interference.

In this paper, we report an intriguing  new observation of
$\lambda/N$ ($N=2$) photonic de Broglie wave interference that has
no classical interpretation and is not associated with the NOON
state. We also show theoretically that photonic de Broglie waves can even
be observed for a pair of photons that are completely separable (i.e., no entanglement in all degrees of freedom) and distinguishable. The experimental and theoretical results suggest that the photonic de Broglie wave is, in fact, not related to the entanglement of the photons, rather it reflects the characteristics (i.e., the indistinguishable pathways) of the measurement scheme.

\section{Experimental observation of photonic de Broglie waves without the NOON state}

\begin{figure}[t]
\includegraphics[width=3.4in]{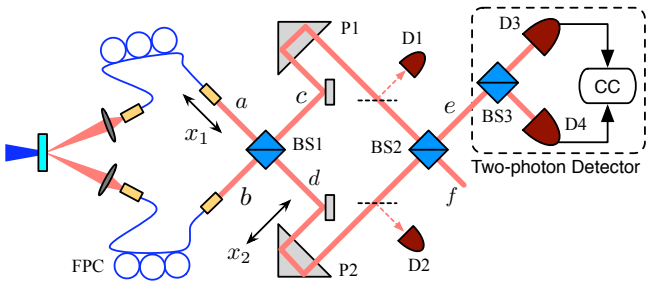}
\caption{\label{setup} Schematic of the experiment. BS1, BS2, and BS3 are 50:50 beam splitters. FPC is the fiber polarization controller and CC is a coincidence counter.}
\end{figure}

Consider the experimental setup shown in Fig.~\ref{setup}. A 405 nm
blue diode laser, with the full width at half maximum (FWHM)
bandwidth of 0.67 nm, pumps a 3 mm thick type-I BBO crystal to
generate, via the SPDC process, a pair of energy-time entangled
photons centered at $\lambda=810$ nm. The photon pair is coupled
into the single-mode optical fiber after passing through the
interference filter with a FWHM bandwidth of 5 nm. For optimal
coupling, the pump was focused at the BBO and the focal spot was
imaged to the single-mode fiber \cite{kwon}.

The photon pair is then sent to a Mach-Zehnder interferometer (MZI),
formed with BS1 and BS2, via the different input ports $a$ and $b$.
The input delay $x_1$ between the photons is controlled by axially
moving the output collimator of a fiber and the interferometer path
length difference $x_2$ is controlled by translating one of the
trombone prisms P2. A two-photon detector, consisting of BS3 and two
single-photon detectors D3 and D4, is placed at the output mode $e$
of MZI for photonic de Broglie wave measurement \cite{eda}. Two
auxiliary detectors D1 and D2 are used to adjust the input delay
$x_1$ by observing the Hong-Ou-Mandel (HOM) interference
\cite{hong}.

\begin{figure}[t]
\includegraphics[width=3.4in]{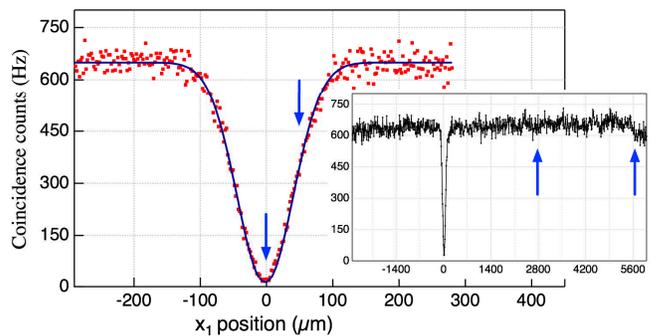}
\caption{\label{hom}  The Hong-Ou-Mandel dip observed with D1 and D2. The dip
visibility is better than 98\%. The arrows represent the $x_1$ positions at which the biphoton interference measurements were performed with the two-photon detector (i.e., coincidences between detectors D3 and D4).}
\end{figure}

First, we consider the well-known photonic de Broglie wave for a
biphoton NOON state and this requires preparing the state
$|\psi\rangle = (|2\rangle_c|0\rangle_d +
|0\rangle_c|2\rangle_d)/\sqrt{2}$ in the MZI \cite{jacob,eda}. This
can be accomplished by using HOM interference: the photon pair
arrives at BS1 (or enters the MZI) simultaneously via the different
input ports $a$ and $b$. The high-visibility HOM interference,
measured in coincidence counts between D1 and D2 as a function of
$x_1$, reported in Fig.~\ref{hom} indicates that when the input
delay is zero, i.e., $x_1=0$, the quantum state of the photons in
the interferometer is indeed the desired biphoton NOON state.

\begin{figure}[t]
\includegraphics[width=3.4in]{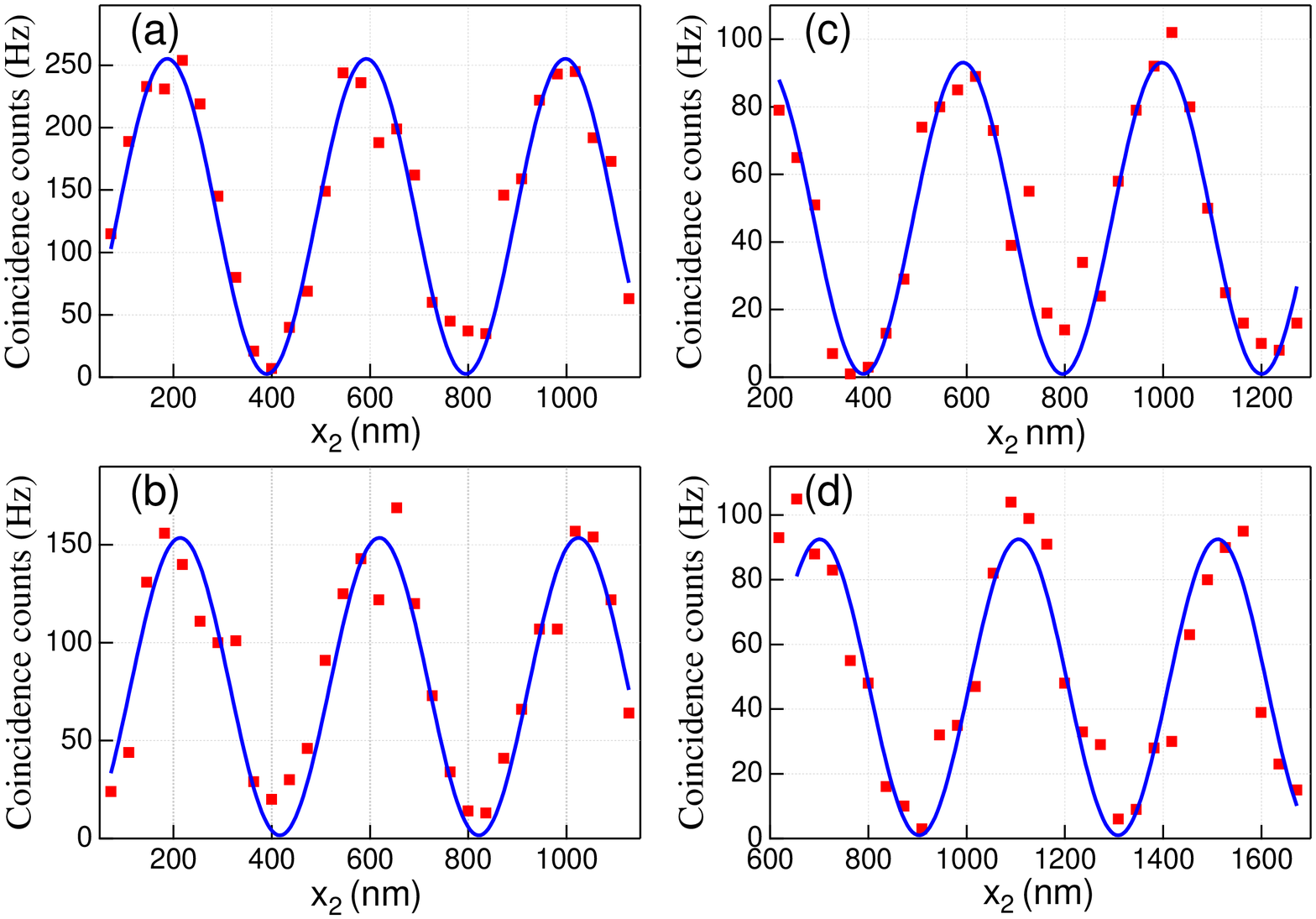}
\caption{\label{mod} Biphoton interference observed at four different $x_1$
positions. (a) $x_1 = 0$ $\mu$m, (b) $x_1 = 62$ $\mu$m, (c) $x_1 =
2.8$ mm. (a) $x_1 = 5.7$ mm. The solid lines are fit to the data
with the modulation wavelength and the visibility fixed at
$\lambda/2=405$ nm and $98$\%, respectively. }
\end{figure}

Observation of the photonic de Broglie wave for the biphoton NOON
state requires i) interfering the biphoton amplitudes
$|2\rangle_c|0\rangle_d$ and $|0\rangle_c|2\rangle_d$ and ii) making
a proper two-photon detection. In the experiment, we set $x_1=0$
with the help of the HOM dip in Fig.~\ref{hom} and the photonic de
Broglie wave corresponding to the biphoton NOON state was observed
at the two-photon detector placed at the output mode $e$ of the MZI.
The result shown in Fig.~\ref{mod}(a) exhibits $\lambda/2$
interference fringes as a function of the MZI path length difference
$x_2$.

We note that the coincidence between single photon detectors placed
at modes $e$ and $f$ also exhibits the interference fringes with the
period $\lambda/2$ \cite{rarity,ou}. This $\lambda/2$ interference
fringe, however, is, not related to the photonic de Broglie wave
since i) the photons are split at BS2 and ii) it may be observed
with the classical coherent state, e.g.,
$|0\rangle_a|\alpha\rangle_b$,  at the input of the MZI
\cite{eda,resch}.

Consider now the situation in which the photons do not enter the MZI
simultaneously. In this case, since the photons do not arrive at BS1
at the same time, HOM interference does not occur and the quantum
state of the photons in the MZI is no longer the biphoton NOON
state. The question we ask is whether the $\lambda/2$ photonic de
Broglie wave would still be observed at the two-photon detector in
mode $e$ (i.e., coincidences between detectors D3 and D4)  in this case.

To probe this question, we intentionally add more time delays in
mode $a$ of the MZI so that $x_1 \neq 0$. The arrows in
Fig.~\ref{hom} indicate the $x_1$ positions at which the biphoton interference measurements  are made with the two-photon detector in mode $e$. First, we set $x_1=62$ $\mu$m and  $x_2$ is scanned for the
two-photon interference measurement. At this $x_1$ location, there
is still some Hong-Ou-Mandel interference as evidenced in Fig.~\ref{hom} (i.e., the coincidence rate is still below the random coincidence rate). The biphoton interference measured with the two-photon detector in this condition is
shown in Fig.~\ref{mod}(b). Interestingly, the observed interference
fringes exhibit the same $\lambda/2$ modulation with no reduction in
visibility. It is intriguing to find that the same high-visibility
interference fringes with $\lambda/2$ modulations are observed even
when $x_1$ is completely out of the Hong-Ou-Mandel dip region. In
Fig.~\ref{mod}(c) and Fig.~\ref{mod}(d), we show the biphoton
interference observed with the two-photon detector at  $x_1=2.8$ mm and at $x_1=5.7$ mm,
respectively. These data  correspond to the $x_1$ positions marked
with the arrows shown in the inset of Fig.~\ref{hom}.

So far, we have established experimentally that the photonic de
Broglie wave can indeed be observed without the NOON state.  (Note
that, differently from Ref.~\cite{resch}, this is a real
second-order quantum effect in the absence of any first-order
interference: the detectors D3 and D4 individually do not show any
modulations.) We now ask whether the shapes of the photonic de
Broglie wave packets would remain the same. This question is probed
by measuring the the photonic de Broglie wave packets for several
different $x_1$ values and the results of these measurements are
shown in Fig.~\ref{packet} \cite{kwon2}.

In Fig.~\ref{packet}(a), we show the typical symmetric Gaussian de
Broglie wave packet for the biphoton NOON state generated by setting
$x_1=0 \mu$m. This case corresponds to Fig.~\ref{mod}(a). For
non-NOON states (i.e., for $x_1 \neq 0$), it is found that the
photonic de Broglie wave packet is modified dramatically. The wave
packet starts to become highly asymmetric (with respect to the
random coincidence rate) as soon as $x_1 \neq 0$, see
Fig.~\ref{packet}(b). The wave packet envelope then takes the shape
of a double-hump and a single-dip for a larger value of $x_1$, see
Fig.~\ref{packet}(c). Eventually, for a sufficiently large $x_1$,
small side peaks starts to appear at $x_2=\pm x_1$, see
Fig.~\ref{packet}(d). Even for very large values of $x_1$, e.g.,
corresponding to the positions marked with arrows in the inset of
Fig.~\ref{hom}, the wave packet shape remains essentially the same
as in Fig.~\ref{packet}(d) but the two side peaks get relocated to
their new positions, $x_2=\pm x_1$~\cite{kwon2}.

\section{Photonic de Broglie wave without the NOON state}

\subsection{Theoretical description}

To explain the observed phenomena theoretically, we start by writing
the monochromatic laser pumped SPDC two-photon state as
\cite{baek08}
\begin{equation}
|\psi\rangle_{\textrm{e}} =  \int d\omega_s d\omega_i \,
\delta(\Delta_{\omega}) \mathrm{sinc}(\Delta_k L/2) e^{i\Delta_k
L/2} |\omega_s,\omega_i \rangle,\label{spdc}
\end{equation}
where the subscripts $i$, $s$, and $p$ refer to the idler, the
signal, and the pump photon, respectively. The thickness of the SPDC
crystal is $L$, $\Delta_{\omega}=\omega_p-\omega_s-\omega_i$, and
$\Delta_k = k_p - k_s-k_i$. Since the pump is a cw diode laser with
a rather large FWHM bandwidth, the SPDC quantum state with cw diode
laser pump should more properly be written as \cite{kwon2}
\begin{equation}
\rho=\int d\omega_p \, \mathcal{S}\left(\omega_p\right) |\psi
\rangle_{\textrm{e}} {}_{\textrm{e}}\langle\psi|,
\end{equation}
where the spectral power density of the pump is assumed to be
Gaussian
\begin{equation}
\mathcal{S}\left(\omega_p\right) \equiv
\exp\left(-{(\omega_p-\omega_{p0})^2}/{2{\Delta\omega_p}^2}\right)/
{\Delta\omega_p\sqrt{2\pi}},
\end{equation}
such that $\int \mathcal{S}(\omega_p) d\omega_p = 1$.

\begin{figure}[t]
\includegraphics[width=3.4in]{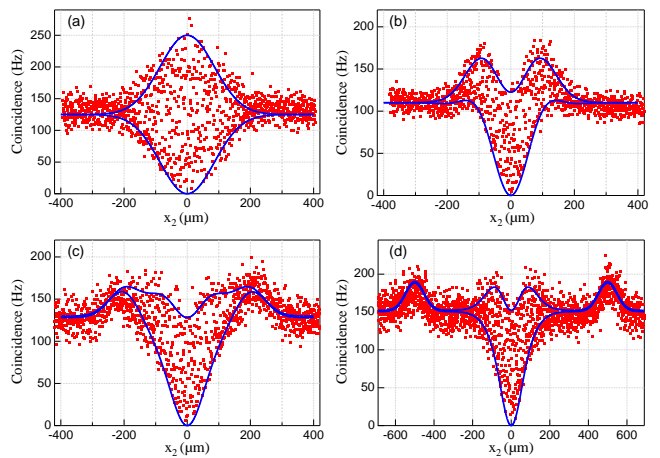}
\caption{\label{packet} The biphoton wave packet measurements with varying input
delays at BS1. (a) $x_1=0$ $\mu$m, (b) $x_1=100$ $\mu$m, (c)
$x_1=200$ $\mu$m, (d) $x_1=500$ $\mu$m. Within the wave packets, the
modulation period is $\lambda/2$ and the visibility around $x_2=0$
is better than 98\%. The solid lines are the wave packet envelopes
calculated using Eq.~(\ref{debroglie}).}
\end{figure}

The HOM interference can be calculated by evaluating
\begin{equation}
R_{12} = \int dt dt' \, tr[\rho \, E_c^{(-)}(t) E_d^{(-)}(t')
E_d^{(+)}(t') E_c^{(+)}(t)],\label{homeq}
\end{equation}
where
\begin{eqnarray}
E_c^{(+)}(t) &=& (i E_a^{(+)}(t-\tau_1) + E_b^{(+)}(t))/\sqrt{2},\\
E_d^{(+)}(t) &=& (E_a^{(+)}(t-\tau_1) + i E_b^{(+)}(t))/\sqrt{2},
\end{eqnarray}
and $\tau_1=x_1/c$. The positive frequency
component of the electric field in mode $a$ is given as
\begin{equation}
E_a^{(+)}(t) = \int d\omega \, a(\omega) \phi(\omega) e^{-i\omega
t},
\end{equation}
 where $a(\omega)$ is the annihilation operator for the signal
photon in mode $a$ and $E_b^{(+)}(t)$ for the idler photon in mode
$b$ is similarly defined. The filter transmission is assumed Gaussian
\begin{equation}
\phi(\omega)=\exp\left(-{(\omega-\omega_{0})^2}/{2{\Delta\omega}^2}
\right) /\sqrt{\Delta\omega\sqrt{\pi}},
\end{equation}
 and $\int |\phi(\omega)|^2
d\omega = 1$. Since the natural bandwidth of SPDC,
$\textrm{sinc}(\Delta_k L/2)$, is much broader than the spectral
filter bandwidth $\Delta\omega$, Eq.~(\ref{homeq}) is calculated to
be
\begin{equation}
R_{12} = 1-\exp(-\Delta\omega^2\tau_1^2/2).\label{Rdip}
\end{equation}
The solid line in Fig.~\ref{hom} is plotted using Eq.~(\ref{Rdip})
with measured spectral filter bandwidth $\Delta\omega$.

For the photonic de Broglie wave measurement, the response of the
two-photon detector in mode $e$ must be considered and it is given
as
\begin{equation}
R_{ee} = \int dt dt' \, tr[\rho \, E_e^{(-)}(t) E_e^{(-)}(t')
E_e^{(+)}(t') E_e^{(+)}(t)],\label{Ree}
\end{equation}
where
\begin{equation}
E_e^{(+)}(t) = (i E_c^{(+)}(t) +E_d^{(+)}(t-\tau_2))/\sqrt{2}
\end{equation}
 and $\tau_2=x_2/c$. Equation (\ref{Ree}) can then be re-written as
\begin{equation}
R_{ee} = \int d\omega_p \mathcal{S}(\omega_p) \int dt dt' \, \left|
\langle 0|E_e^{(+)}(t') E_e^{(+)}(t)|\psi
\rangle_{\textrm{e}}\right|^2,\label{Ree2}
\end{equation}
where $\langle0|$ denotes the vacuum state. The biphoton amplitude $\langle 0|E_e^{(+)}(t') E_e^{(+)}(t)|\psi
\rangle_{\textrm{e}}$ contains important information about the quantum interference and, when expanded using the electric field operators at input modes $a$ and $b$, is calculated to be
\begin{widetext}
\begin{equation}
\begin{array}{ll}
 \langle0|E_e^{(+)}(t') E_e^{(+)}(t)|\psi
\rangle_{\textrm{e}}=\\
 \frac{i}{4}\langle 0|
\left[ \begin{array}{l}   E_a^{(+)}(t-\tau_1-\tau_2)
E_b^{(+)}(t'-\tau_2) + E_a^{(+)}(t'-\tau_1-\tau_2)
E_b^{(+)}(t-\tau_2)\\- E_a^{(+)}(t-\tau_1) E_b^{(+)}(t')-
E_a^{(+)}(t'-\tau_1) E_b^{(+)}(t) \\- E_a^{(+)}(t-\tau_1)
E_b^{(+)}(t'-\tau_2)- E_a^{(+)}(t'-\tau_1) E_b^{(+)}(t-\tau_2)\\+
E_a^{(+)}(t-\tau_1-\tau_2)
E_b^{(+)}(t') + E_a^{(+)}(t'-\tau_1-\tau_2) E_b^{(+)}(t)\\
\end{array}\right]|\psi
\rangle_{\textrm{e}}
 .\end{array}
\\
\label{amplitudes_entangled}
\end{equation}
\end{widetext}
Note that only non-zero biphoton amplitudes are written in the above equation: terms that contain $E_a^{(+)}E_a^{(+)}$ and $E_b^{(+)}E_b^{(+)}$ are eventually calculated to be zero because of the nature of the input state $|\psi\rangle_{\textrm{e}}$.

If we now consider the two-photon detector shown in Fig.~\ref{setup}, the
normalized coincidence rate between D3 and D4 corresponds to
$R_{ee}$ and is given as
\begin{eqnarray}
R_{34} &=&  \frac{1}{4} \{4+\exp(-(\tau_1-\tau_2)^2\Delta\omega^2/2)\nonumber\\
            &+& \exp(-(\tau_1+\tau_2)^2 \Delta\omega^2/2)
           - 2 \exp(-\tau_2^2\Delta\omega^2/2)\nonumber\\
           &-& 2\cos{(2\omega_{0}\tau_2)}\exp(-\tau_2^2\Delta\omega_{e}^2/2)\nonumber\\
           &&\times (1+\exp(-\tau_1^2\Delta\omega^2/2))\},\label{debroglie}
\end{eqnarray}
where $1/{\Delta\omega_{e}^2} \equiv 1/\Delta\omega_{p}^2 +
1/\Delta\omega^2$.

Equation (\ref{debroglie}) clearly shows that the  $2\omega_0$ or
$\lambda_0/2$ interference fringe (corresponding to the photonic de
Broglie wavelength), in fact, is not related to the biphoton NOON
state condition $\tau_1=0$. As long as $\tau_2$ is within the
effective coherence length $\sqrt{2}/\Delta \omega_e$, the biphoton
photonic de Broglie wave interference can be observed regardless of
the $\tau_1$ value.

Another interesting feature of Eq.~(\ref{debroglie}) is that the
shape of the biphoton de Broglie wave packet is $\tau_1$ dependent
while the period of interference fringes remains the same at
$2\omega_0$. Note also that the maximum interference visibility is
not affected by $\tau_1$. The theoretical result in Eq.~(\ref{debroglie}) is found to be in excellent agreement  with the experimental data in Fig.~\ref{mod}.

\subsection{The Feynman diagram}

The interesting features of the
biphoton de Broglie interference in this experiment can be
intuitively understood by analyzing the Feynman diagrams
representing the two-photon detection amplitudes.

\begin{figure}[t]
\includegraphics[width=3.4in]{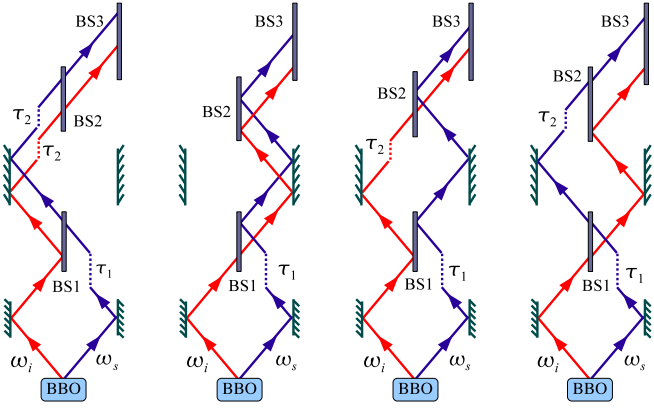}
\caption{\label{feynman} The Feynman paths for the photon pair. All the Feynman paths become
indistinguishable when $\tau_2=0$, regardless of $\tau_1$ values. Note that each line (top to bottom) in Eq.~(\ref{amplitudes_entangled}) corresponds to each Feynman path (left to right).}
\end{figure}

Given the experimental setup in Fig.~\ref{setup}, there exist four Feynman
paths in which the photon pair exits BS2 via the output mode $e$ and these Feynman paths are shown in Fig.~\ref{feynman}. Since the signal, $\omega_s$, and idler, $\omega_i$, photons must always transmit/reflect or
reflect/transmit at BS3 to contribute to a final two-photon
detection event, each Feynman path shown in Fig.~\ref{feynman}
branches off into two final Feynman amplitudes. There are, thus,
total of eight Feynman paths which lead to a detection event at the two-photon detector in Fig.~\ref{setup}.

The photonic de Broglie wavelength observed in
Fig.~\ref{mod} is a manifestation of quantum interference among
these Feynman paths. For arbitrary $\tau_1$ and $\tau_2$, the
Feynman paths shown in Fig.~\ref{feynman} are clearly
distinguishable (in time). However, if $\tau_2=0$, all
Feynman paths become indistinguishable, regardless of $\tau_1$
values. This is confirmed theoretically in Eq.~(\ref{debroglie}) and
experimentally in Fig.~\ref{mod}: high-visibility $2\omega_0$ or
$\lambda/2$ interference fringes are observed when $\tau_2$ is
scanned around $\tau_2=0$.

In addition, it is shown in Fig.~\ref{packet} that the shape of the biphoton wave packet is dependent
on the $\tau_1$ value. In the case that $\tau_1=0$, the third and
fourth Feynman paths in Fig.~\ref{feynman} cancel out and the wave
packet envelope is determined by the overlap between the first two
Feynman paths. As shown in Fig.~\ref{packet}(a), the result is a
Gaussian wave packet whose width is determined by $\Delta\omega_e$.

Consider now the case of $\tau_2=\tau_1$. The signal and idler photons arrive
simultaneously at BS2 for the third Feynman path in
Fig.~\ref{feynman} and, because of the Hong-Ou-Mandel effect, the two photons will always exit BS2 via the same output port. If we now consider the case of $\tau_2=-\tau_1$, the same situation occurs for the fourth Feynman path in Fig.~\ref{feynman}. Therefore the detection probability of the third and the fourth
Feynman paths would increase  twice as big compared to $\tau_2\neq\pm\tau_1$. The net results are the distinct side peaks observed at  $\tau_2=\pm\tau_1$ in Fig.~\ref{packet}(d).

In general, i.e., $\tau_1 \neq 0$, all the eight Feynman paths contribute to quantum interference in a complex manner so an intuitive explanation becomes difficult.

\section{Photonic de Broglie wave interference without entanglement }

So far, we have shown experimentally and theoretically that  the photonic de
Broglie wave is in fact not related to the photon number-path
entangled  or the NOON state. The photonic de Broglie wave, instead, appears to be linked to the underlying spectral entanglement of SPDC photons which are used for both experimental observation and theoretical analysis
\cite{baek08}. The question then becomes whether the two input
photons need to have any entanglement at all to exhibit the photonic
de Broglie wave phenomenon.

\subsection{Photonic de Broglie wave interference for two identical photons with no entanglement}

Consider two single-photons with identical spectra and polarization, each emitted from a separate single-photon source. It is known that HOM interference can occur with a pair of identical single-photons \cite{san,mos}. The biphoton NOON state resulting from HOM interference should then exhibit the photonic de Broglie wave.

The relevant question therefore is what would happen when there is no HOM
interference between the two identical single-photons with  no \textit{a priori} entanglement. Would the photonic de Broglie wave still be observed in the absence of any entanglement between the photons?

To investigate this question, let us consider a single-photon in the pure state at each input mode of the MZI in Fig.~\ref{setup}. Since the joint quantum state of the two single-photons at the input modes of the MZI is separable, it can be written as
\begin{equation}
|\psi\rangle_{\textrm{s}} =  \int d\omega_a \ \varphi(\omega_a)|\omega_a\rangle \otimes \int d\omega_b \ \varphi(\omega_b) |\omega_b \rangle,\label{sep}
\end{equation}
where the single-photon spectral amplitude is assumed to be Gaussian
\begin{equation}
\varphi(\omega)=\exp\left(-{(\omega-\omega_{0})^2}/{2{\Delta\omega}^2} \right) /\sqrt{\Delta\omega\sqrt{\pi}},
\end{equation}
and $\int |\varphi(\omega)|^2 d\omega = 1$.

Given the input quantum state as in Eq.~(\ref{sep}), the response of
the MZI can now be studied. First, the single-photon detection rates
at D3 and D4 are calculated to be constant, completely independent of
$x_1$ and $x_2$. This is because the single-photon detection
probabilities due to the single-photons in mode $a$ and  in mode $b$
have the same Gaussian envelopes but  are out of phase by
$180^\circ$. In other words, similarly to the case of entangled-photon
pairs at the input, no first-order interference can be observed.
Second, the two-photon detection rates for the photonic de Broglie
wave measurement can be calculated
by evaluating
\begin{equation}
R_{ee}^{(s)} = \int dt dt' \, tr[\rho^{(s)} \, E_e^{(-)}(t)
E_e^{(-)}(t') E_e^{(+)}(t') E_e^{(+)}(t)],\label{Reesep}
\end{equation}
where $\rho^{(s)}=|\psi\rangle_{\textrm{s}} {}_{\textrm{s}}\langle
\psi|$. The above equation can then be re-written as
\begin{equation}
R_{ee}^{(s)} =  \int dt dt' \, \left| \langle 0|E_e^{(+)}(t') E_e^{(+)}(t)|\psi
\rangle_{\textrm{s}}\right|^2.\label{Ree3}
\end{equation}
The biphoton amplitude $\langle 0|E_e^{(+)}(t') E_e^{(+)}(t)|\psi \rangle_{\textrm{s}}$ in Eq.~(\ref{Ree3}) is evaluated to be
\begin{widetext}
\begin{equation}
\begin{array}{ll}
 \langle0|E_e^{(+)}(t') E_e^{(+)}(t)|\psi
\rangle_{\textrm{s}}=\\
 \frac{i}{4}\langle 0|
\left[ \begin{array}{l}   E_a^{(+)}(t-\tau_1-\tau_2)
E_b^{(+)}(t'-\tau_2) + E_a^{(+)}(t'-\tau_1-\tau_2)
E_b^{(+)}(t-\tau_2)\\- E_a^{(+)}(t-\tau_1) E_b^{(+)}(t')-
E_a^{(+)}(t'-\tau_1) E_b^{(+)}(t) \\- E_a^{(+)}(t-\tau_1)
E_b^{(+)}(t'-\tau_2)- E_a^{(+)}(t'-\tau_1) E_b^{(+)}(t-\tau_2)\\+
E_a^{(+)}(t-\tau_1-\tau_2)
E_b^{(+)}(t') + E_a^{(+)}(t'-\tau_1-\tau_2) E_b^{(+)}(t)\\
\end{array}\right]|\psi
\rangle_{\textrm{s}}
 .\end{array}
\\
\label{amplitudes_separable}
\end{equation}
\end{widetext}

Finally, the normalized coincidence rate on D3 and D4 in Fig. \ref{setup} is
proportional to $R_{ee}^{(s)}$ and is given as
\begin{eqnarray}
R_{34}^{(s)} &=&  \frac{1}{4} \{4+\exp(-(\tau_1-\tau_2)^2\Delta\omega^2/2)\nonumber\\
            &+& \exp(-(\tau_1+\tau_2)^2 \Delta\omega^2/2)
           - 2 \exp(-\tau_2^2\Delta\omega^2/2)\nonumber\\
           &-& 2\cos{(2\omega_{0}\tau_2)}\exp(-\tau_2^2\Delta\omega^2/2)\nonumber\\
           &&\times (1+\exp(-\tau_1^2\Delta\omega^2/2))\}.\label{debrogliesingle}
\end{eqnarray}

\begin{figure*}[t]
\includegraphics[width=5in]{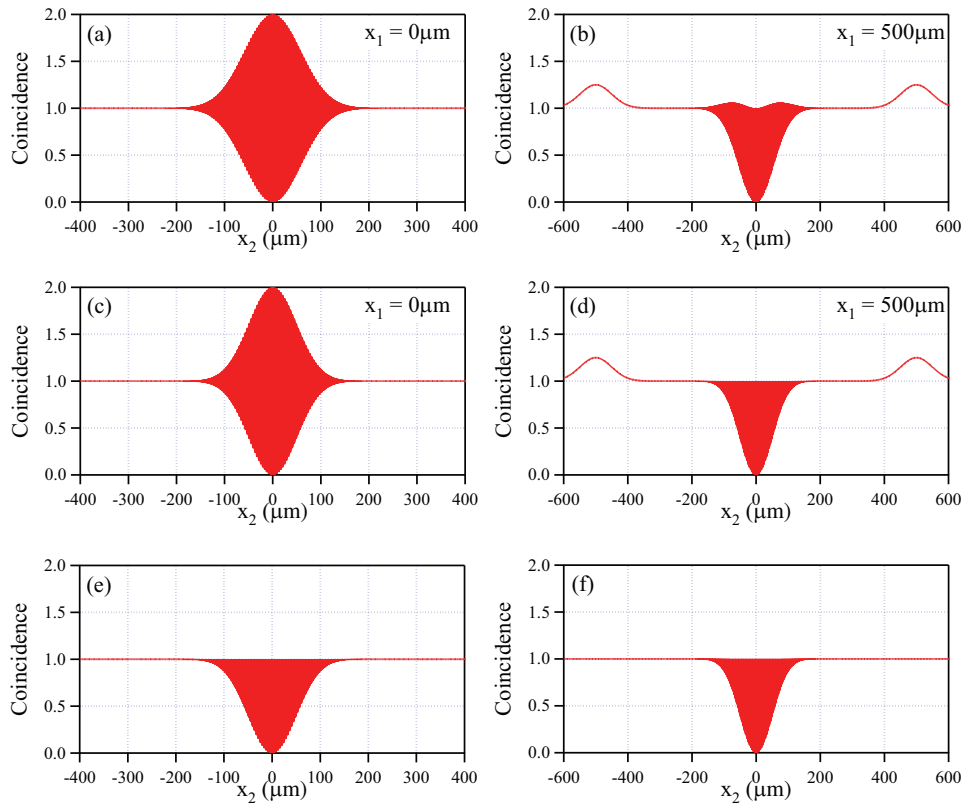}
\caption{\label{theory} Calculated photonic de Broglie wave packets for SPDC
photons, (a) and (b); for two identical single-photons with no entanglement, (c) and (d); and for two
distinguishable (orthogonally polarized) single-photons with no entanglement, (e) and
(f). The plots (a) and (b), (c) and (d), and (e) and (f) are due to the theoretical results in Eq.~(\ref{debroglie}), Eq.~(\ref{debrogliesingle}), and Eq.~(\ref{debrogliedist}), respectively. For SPDC photons, the pump bandwidth $\Delta\omega_p$ is assumed to be 2 nm FWHM and the signal and the idler photons are
filtered with 5 nm FWHM filters. For single-photons, they are assumed to
have FWHM bandwidth of 5 nm. Note that, since Eq.~(\ref{debrogliedist}) is $\tau_1$ independent, (e) and (f) are identical plots with different ranges.  }
\end{figure*}

It is interesting to note that the result in Eq.~(\ref{debrogliesingle}) is identical to
Eq.~(\ref{debroglie}) but with $\Delta\omega_e$ replaced by
$\Delta\omega$. Effectively, this means that SPDC pumped with a very broadband pump
laser would give the identical result as that of two separable
single-photon states. The theoretical results summarized in
Fig.~\ref{theory} show that the separable two-photon state of
Eq.~(\ref{sep}) at the input of the MZI gives nearly the same result
as that of SPDC photons pumped with a laser with 2 nm FWHM bandwidth
for both the NOON state ($x_1=0$ $\mu$m) and non-NOON state ($x_1 \neq
0$ $\mu$m) conditions.

\subsection{Photonic de Broglie wave interference for two distinguishable (orthogonally polarized) photons with no entanglement}

In the previous section, we have seen that entanglement is in fact not necessary for observing the  photonic de Broglie wave interference of two photons. It was however assumed that the two input single-photons were identical. In this section, we discuss the general case in which the two input single-photons are orthogonally polarized so that they are completely distinguishable. Note that the experimental schematic is kept the same as in Fig.~\ref{setup}: no polarization-information erasing polarizers are added to the setup.

For two orthogonally polarized single-photons, the joint quantum state is written as
\begin{equation}
|\psi\rangle_{\textrm{dist}} =  \int d\omega_a \
\phi(\omega_a)|\omega_a^{H}\rangle \otimes \int d\omega_b \
\phi(\omega_b) |\omega_b^{V} \rangle,\label{dist}
\end{equation}
where the superscripts $H$ and $V$ refer to horizontal and vertical polarization states, respectively. The counting rate at the two-photon detector, see Fig.~\ref{setup}, in the output mode $e$ of BS2 is then given as
\begin{widetext}
\begin{equation}
R_{ee}^{(dist)} = \sum_{p_1,p_2\in \{H,V\}} \int dt dt' \,
tr[\rho^{(dist)} \, E_e^{p_1(-)}(t) E_e^{p_2(-)}(t')
E_e^{p_2(+)}(t') E_e^{p_1(+)}(t)],\label{Reedist}
\end{equation}
where superscripts $p_1$ and $p_2$ denote polarizations
 and $\rho^{(dist)}=|\psi\rangle_{\textrm{dist}} \
{}_{\textrm{dist}}\langle \psi|$. Equation (\ref{Reedist}) can then be re-written as
\begin{eqnarray}
R_{ee}^{(dist)}&=&\int dt dt' \, \sum_{p_1,p_2\in \{H,V\}} \left|
\langle 0|E_e^{p_2(+)}(t')
E_e^{p_1(+)}(t)|\psi \rangle_{\textrm{dist}}\right|^2\\
&=&\int dt dt'  \left( \left| \langle 0|E_e^{H(+)}(t') E_e^{V(+)}(t)|\psi
\rangle_{\textrm{dist}}\right|^2+\left| \langle 0|E_e^{V(+)}(t')
E_e^{H(+)}(t)|\psi \rangle_{\textrm{dist}}\right|^2\right).\nonumber
\label{Reedist_form_change}
\end{eqnarray}
\end{widetext}
Note that terms that include electric field operators $E_e^{H(+)}E_e^{H(+)}$ and
$E_e^{V(+)}E_e^{V(+)}$ are not shown because they eventually are calculated to be zero since the input photons are orthogonally polarized.

The biphoton amplitudes are then expanded as
\begin{widetext}
\begin{equation}
\begin{array}{ll}
 \langle0|E_e^{H(+)}(t') E_e^{V(+)}(t)|\psi
\rangle_{\textrm{dist}}=\\
 \frac{i}{4}\langle 0|
\left[ \begin{array}{l}   E_a^{H(+)}(t-\tau_1-\tau_2)
E_b^{V(+)}(t'-\tau_2) - E_a^{H(+)}(t-\tau_1) E_b^{V(+)}(t')\\-
E_a^{H(+)}(t-\tau_1) E_b^{V(+)}(t'-\tau_2)+
E_a^{H(+)}(t-\tau_1-\tau_2)
E_b^{V(+)}(t') \\
\end{array}\right]|\psi
\rangle_{\textrm{dist}}
 ,\end{array}
\\
\label{amplitudes1}
\end{equation}
and
\begin{equation}
\begin{array}{ll}
 \langle0|E_e^{V(+)}(t') E_e^{H(+)}(t)|\psi
\rangle_{\textrm{dist}}=\\
 \frac{i}{4}\langle 0|
\left[ \begin{array}{l}   E_a^{H(+)}(t'-\tau_1-\tau_2)
E_b^{V(+)}(t-\tau_2) - E_a^{H(+)}(t'-\tau_1) E_b^{V(+)}(t)\\-
E_a^{H(+)}(t'-\tau_1) E_b^{V(+)}(t-\tau_2)+
E_a^{H(+)}(t'-\tau_1-\tau_2)
E_b^{V(+)}(t) \\
\end{array}\right]|\psi
\rangle_{\textrm{dist
}}
 .\end{array}
\\
\label{amplitudes2}
\end{equation}
Finally, the normalized output of the two-photon detector (i.e., coincidence between D3 and D4) is calculated to be,
\begin{eqnarray}
R_{34}^{(dist)} &=&  \frac{1}{4} \{4- 2
\exp(-\tau_2^2\Delta\omega^2/2)-
2\cos{(2\omega_{0}\tau_2)}\exp(-\tau_2^2\Delta\omega^2/2)\}
           .\label{debrogliedist}
\end{eqnarray}
\end{widetext}

It is interesting to note that Eq.~(\ref{debrogliedist}) also shows $2\omega_0$ modulation as in the case of two identical single-photons,  Eq.~(\ref{debrogliesingle}), and as in the case of a pair of SPDC photons, Eq.~(\ref{debroglie}). This result, therefore, reveals that photonic de Broglie wave interference is not only unrelated to the NOON state, but it can also be observed with completely unentangled and distinguishable photons. Note also that  Eq.~(\ref{debrogliedist}) is completely independent of $\tau_1$ and Eq.~(\ref{debrogliedist}) can actually be obtained from Eq.~(\ref{debrogliesingle}) by letting $\tau_1 \rightarrow \infty $.

Equation (\ref{debrogliedist}) is plotted in Fig.~\ref{theory}(e) and Fig.~\ref{theory}(f). The plots show very clearly that high-visibility photonic de Broglie wave interference appear for two orthogonally polarized single-photons. Note, however, that the shape of the wave packet in Fig.~\ref{theory}(e) is quite different from Fig.~\ref{theory}(a) and Fig.~\ref{theory}(c) but rather similar to Fig.~\ref{theory}(b) and Fig.~\ref{theory}(d). This comes from the fact that Eq.~(\ref{debrogliedist}) is $\tau_1$ independent and the other two results converge toward Eq.~(\ref{debrogliedist}) as $\tau_1$ gets bigger. This fact is also reflected in the absence of side peaks in Fig.~\ref{theory}(f).


\section{Conclusion}

It is interesting to discuss the connection between the photonic de Broglie wave interference and entanglement between the input photons. The monochromatic-pumped SPDC in Eq.~(\ref{spdc}) is strongly
energy-time entangled and, as the pump bandwidth is increased, the
degree of energy-time entanglement is reduced \cite{kim05}. The experimental and theoretical results on photonic de Broglie wave interference for broadband-pumped SPDC shown in Fig.~\ref{packet}(d) and in
Fig.~\ref{theory}(b) make it clear that the quality of the photonic de Broglie wave interference for non-NOON states is not affected by the reduced energy-time entanglement between the photon pair. Furthermore, Fig.~\ref{theory}(d) and Fig.~\ref{theory}(f) show that even two unentangled and distinguishable (orthogonally polarized) single-photons lead to essentially the same photonic de Broglie wave interference.

These results therefore reveal that entanglement between the two photons plays essentially no role in the manifestation of the photonic de Broglie wave interference. Rather, it is the measurement scheme (i.e., indistinguishable pathways established by the measurement scheme) that brings out the photonic de Broglie wave phenomenon \cite{kim05b}.

The  experimental and theoretical results in this paper apply to $N=2$ photonic de Broglie wave interference. We, however, believe that it should be possible to extend the conclusions to the $N$ photon case.

\section*{Acknowledgements}

This work was supported, in part, by the National Research Foundation of Korea (KRF-2006-312-C00551, 2009-0070668, and 2009-0084473) and the Ministry of Knowledge and Economy of Korea through the Ultrafast Quantum Beam Facility Program.



\end{document}